\documentstyle[amssymb,epsf,amsmath,amsfonts,twocolumn,prl,aps]{revtex}

\tighten

\begin{document}
\draft

\twocolumn[\hsize\textwidth\columnwidth\hsize\csname
@twocolumnfalse\endcsname

\title{Perturbative and Numerical Analysis of Tilted Cosmological Models of Bianchi type V}

\author{Michael Bradley and Daniel Eriksson}
\address{Department of Physics, Ume{\aa}
University, Ume{\aa}, Sweden \\
e-mail: michael.bradley@physics.umu.se, daniel.eriksson@physics.umu.se}

\date{\today}
\maketitle

\begin{abstract}
  Cosmological models of Bianchi type V and I containing a perfect fluid with a linear 
  equation of state plus cosmological constant are investigated using a tetrad approach where our variables 
  are the Riemann tensor, the Ricci rotation coefficients and a subset of the tetrad 
  vector components. This set, in the following called $S$, describes a spacetime when its
  elements are constrained by certain integrability conditions and due to a theorem by Cartan this set 
  gives a complete local 
  description of the spacetime. 
  The system obtained by imposing the integrability conditions and Einstein's equations can be reduced
  to an integrable system of five   
  coupled first order ordinary differential equations. The general solution is tilted and describes a fluid with  
  expansion, shear and vorticity. 
  With the help of standard bases for Bianchi V and I the full line element is 
  found in terms of the elements in $S$.
 We then construct the solutions to the linearized equations
  around the open Friedmann model. 
 The full system is also
  studied numerically and the perturbative solutions agree well with the numerical ones in
  the appropriate domains. We also give some examples of numerical solutions in the
  non-perturbative regime. 
\end{abstract}

\pacs{PACS:numbers: 04.20.-q, 98.80.-k}

\vspace{2ex}
]

\section{\smallskip Introduction}  \label{intro}

In this paper general Bianchi types V and I 
perfect fluids with linear equation of state and cosmological
constant are studied.
In general these spacetimes are tilted and in particular there are solutions with rotating matter.
It has been difficult to find exact solutions with both expansion and nonzero rotation of the matter flow.
To our knowledge the only known exact homogeneous perfect fluid solution with 
rotation and expansion is the self-similar radiation-filled Bianchi type $\rm{VI}_0$ found by Rosquist \cite{Rosquist}. 

A number of rotating imperfect fluid solutions with heatflow are known, see e.g. \cite{Novello}. 
Since for rotating matter the hypersurfaces of homogeneity are tilted with respect to the restframe 
of matter, local space will not look homogeneous. Hence heatflow is expected and for some solutions
the heatflow can be related to a temperature gradient \cite{Sviestins}, but often with unrealistic coefficient of 
conductivity.  Normally the heat conductivity is negligibly small and a perfect fluid approximation should work well. 

For treatments of the properties of homogeneous rotating models in general
see \cite{KingEllis,Peresetskii} and for some perturbative calculations see 
\cite{DemiGris,BayCoop,BatCoh}. In \cite{CollinsEllis} the qualitative behaviour of locally rotationally 
symmetric (LRS) Bianchi V solutions is analysed and in particular expressions for different quantities
at early and late times are given. 
There are a number of works on tilted Bianchi cosmologies using a dynamical system approach.
The irrotational subcase of type V was studied in \cite{HewittWainwright}. 
Recently the late time behaviour of tilted Bianchi models including type V was considered in
\cite{ColeyHervik}. The stability of non-tilted Bianchi models against tilt was studied in \cite{BarrowHervik}.

To find the solutions we use a method for construction of solutions to Einstein's equations,
\cite{Brans,KarlLind,BradKarl,BradMark},
based on the invariant classification scheme by Cartan and Karlhede \cite{Cartan,Karl}. The method
is shortly described in section \ref{sec:construct}. In section \ref{sec:V} the method is
applied to Bianchi V and I models. First choice of frame and the set (called $S$)
of quantities needed to specify the spacetime are given. Then the structure of the 
isometry group is imposed, giving relations
among the elements in $S$. Next the integrability conditions for the set $S$ to describe
a geometry are imposed together with Einstein's equations. 
The general system is reduced to an integrable system of five coupled first order ordinary differential
equations. With the help of standard bases for Bianchi V and I the full line-element is
found up to quadratures in terms of the elements in $S$.
The subclass of orthogonal solutions is easily solved, but all these solutions are well known.
For LRS dust the equations can also be integrated \cite{Farnsworth,MarkBrad}.

In section \ref{sec:pert}
we consider first order perturbations around the open Friedmann universe. The
general first order solution depends on five constants of integration, the same number as for
the general exact solution, and has nonzero expansion, rotation and shear. 
Finally, in section \ref{sec:numerical} the general system is studied numerically and the results
agree well with the perturbative calculations in the allowed range. Examples of numerical
solutions in the non-perturbative regime are given.

\section{Construction of solutions to Einstein's equations in terms of curvature invariants}\label{sec:construct}

A brief summary of the method is given here. For more details see \cite{BradMark,Marklund}.

According to a theorem by Cartan spacetimes are locally completely determined by
a set consisting of the components of the Riemann tensor and a finite number of its covariant derivatives
in a frame with constant metric $\eta_{ij}$ 
\cite{Cartan,Karl}. This set will henceforth be called $R^{p+1}=\{R_{ijkl},R_{ijkl;m_1},....R_{ijkl;m_1....m_{p+1}}\}$, 
where $p$ is such that the components
in $R^{p+1}$ are functionally dependent on those in $R^p$ (as functions
of both the coordinates $x^{\mu}$ on the manifold and the parameters of the Lorentz
group $\xi^{\Upsilon}$). 
This description can be used to classify geometries and also to construct new solutions
to Einstein's equations in terms of the set $R^{p+1}$.

Assume that we have symmetries such that $R^{p+1}$ only 
depends on $x^{\alpha}$, 
$ \alpha = 1,2,..., l$ $<n=$dimension of spacetime, (in some canonical coordinates)
 and rotations in the
$ab$-planes , $\{ {{}^{a}}_{b} \} = 1,...,m<n(n-1)/2$  
(with frames adopted to the rotational symmetries). Here $l=n-
\mathrm{dim(orbits)}$ and $m=n(n-1)/2-\mathrm{dim(isotropy  \,\, group)}$. 
The set $R^{p+1}$ is completely determined by the smaller
set  $S=\{ {R^p}_{qkl}, {\gamma^a}_{bi}, {x^{\alpha}}_{|i} \}$  
where $x^{\alpha}_{|i}\equiv X_i(x^{\alpha})=X_i^{\,\ \mu}\partial x^{\alpha}/\partial x^{\mu}$ 
are the derivatives with respect to the frame vectors and ${\gamma^i}_{jk}$ are the Ricci rotation
coefficients. Here the numbering is such that $\{ {{}^{p}}_{q} \} = m+1,...,n(n-1)/2$ are the 
complementary rotations to the $\{{{}^{a}}_{b}\}$ ones, i.e., those that keep the
set $R^{p+1}$ unchanged. 

A set $R^{p+1}$ together with a constant frame metric $\eta_{ij}$ describes a geometry iff certain integrability conditions,
being equivalent to the Ricci identities (including the commutators for 
the essential coordinates) 
and part of the Bianchi identities,
are satisfied. In practice it is often easier to work 
in a fixed frame than to let the components in $R^{p+1}$ depend explicitly on 
the parameters ${\xi}^{\Upsilon}$ of the orthogonal group. The Ricci identies then
split into the commutators for the essential coordinates $x^{\alpha}$
\begin{equation}
  {x^{\alpha}}_{[|i,\beta}{x^{\beta}}_{| j]} + {x^{\alpha}}_{|m}
  {{\gamma}^{m}}_{[ij]} = 0,
\label{eq:d2x}
 \end{equation}
and the Riemann equations for rotations in the $ab$-planes
 \begin{equation}
  {R^{a}}_{bij} = 2{{\gamma}^{a}}_{b[j,\alpha }{x^{\alpha}}_{|i]} +
  2{{\gamma}^{ak}}_{[j}{{\gamma}}_{bki]} +
  2{{\gamma}^{a}}_{bk}{{\gamma}^{k}}_{[ij]}
\label{eq:Rie}
 \end{equation} 
(the antisymmetrizations are only over $ij$). When using the description in
terms of the smaller set $S$
equations (\ref{eq:Rie}) merely serve to define the ${R^{a}}_{bij}$-components of the
Riemann tensor.
Note that equations (\ref{eq:Rie}) give 36 equations in the case without isotropies and
the restriction to 20 independent components come from the cyclic identity below 
or equations (\ref{eq:d2x}).
Since not all commutors or Riemann equations are used when the spacetime
has symmetries some more integrability conditions are needed. They are parts
of the cyclic and Bianchi identities
 \begin{eqnarray}
  {R^{t}}_{[ijk]}   = & 0 &, \,\,\, t = l+1,...,n \, , \label{eq:cycl} \\
  {R^{p}}_{q[ij;k]} = & 0 &, \,\,\, \{ {{}^{p}}_{q} \} =
  m+1,...,n(n-1)/2 \, ,
\label{eq:Bianchi}
 \end{eqnarray}
where $t=l+1,...,n$ numbers the symmetry directions 
(in a suitable numbering of the frame vectors).

The above description in terms of $R^{p+1}$ can be used to find new solutions
to Einstein's equations. This approach has the 
advantage of being coordinate invariant. Also, since the components in $R^{p+1}$ 
correspond to directly measurable quantities, physical requirements
are easy to impose. Another advantage is that the set of equations
(\ref{eq:d2x}),  (\ref{eq:Rie}), (\ref{eq:cycl}) and (\ref{eq:Bianchi})
is a subset of the ones used in different tetrad methods like the ones in, e.g.,
\cite{EllisMacCallum} or the Newman-Penrose method \cite{NewmanPenrose}.
The procedure is:
 \begin{enumerate}
  \item Choose a set $R^{p+1}$  (or equivalently $S$).
   Some of the elements in $R^{p}$ or functions of them 
    are chosen as coordinates. Einstein's
    equations are imposed by restricting the Ricci tensor. 
  \item Impose the integrability conditions (\ref{eq:d2x}),  (\ref{eq:Rie}), (\ref{eq:cycl}) and
    (\ref{eq:Bianchi}), leading to a set of first order differential
    equations (together with algebraic constraints) for the elements in
    $R^{p+1}$ ($S)$.
 \end{enumerate}
Solving this set of equations gives $R^{p+1}$ and hence a complete local description
of the geometry. 
If the geometry does not have any symmetries, one can
solve for all the 1-forms from
$dx^{\mu}=x^{\mu}_{\,\, \vert i}\omega^i$
and hence get the 
full line-element $d^2 s =
{\eta}_{ij} {\omega}^{i}  {\omega}^{j}$.

When there are symmetries one only obtains part of the 1-forms, but one
may determine the Lie-algebra of the isometry group, \cite{KarlMac}, and from this it is often
possible to make an ansatz for the remaining 1-forms. 
Cartan's equations look the same in $F(M)$, the bundle of
frames on $M$, as in $M$
\begin{eqnarray}\label{Cartan1}
   d\, \omega^i     &   =   & \omega^j \wedge \omega^i_{\; j}          \\ \label{Cartan2}
    d\, \omega^i_{\; j}    &  =   &   -\omega^i_{\;\; k} \wedge \omega^k_{\;\; j} + 
\tfrac{1}{2}  R^i_{\; jkl}\omega^k \wedge \omega^l     
\end{eqnarray}
but now $d\, =d\, _x+d\, _{\xi}$. The connection forms $\omega^i_{\; j}$ will now be linearly 
independent of the 1-forms $\omega^i$, and can be written as
\begin{equation}\label{connect}
\omega^i_{\; j}=\gamma^i_{\; jk}\omega^k + \tau^i_{\; j} \equiv 
\gamma^i_{\; jk}\omega^k + \tau^i_{\; j \Upsilon}{\rm d}\, \xi^{\Upsilon}
\end{equation}
where $\tau^i_{\; j}$ are the generators of the orthogonal group.
With the same notation for indices as above Cartan's equations split into those corresponding
to essential coordinates and rotations ($x^{\alpha}$ and  $\tau^a_{\,\, b}$) and those corresponding
to symmetry coordinates and rotations ($x^t$ and $\tau^p_{\,\, q}$)
\begin{equation}\label{eq:Cartan4}
   d\, \omega^t  =  \omega^j \wedge \omega^t_{\; j}  , \,\,\,
    d\, \omega^p_{\; q}   =   -\omega^p_{\;\; k} \wedge \omega^k_{\;\; q} + 
\tfrac{1}{2}  R^p_{\; qkl}\omega^k \wedge \omega^l     \, .
\end{equation}
The 1-forms that are missing can be found from these equations. 
Since spacetime locally is completely determined by $R^{p+1}$, it is sufficient to find 
\emph{one} linearly independent solution
to equations (\ref{eq:Cartan4}). 
All other solutions will be locally equivalent to this one up to coordinate
transformations. 
When projecting onto the orbits, given by $dx^{\alpha}=0$ and $\tau^a_{\; b}=0$,  equations (\ref{eq:Cartan4})
reduce to the Lie-algebra of the isometry group (since the group acts simply transitive 
on the orbits in $F(M)$ \cite{RyanShep}) and the structure constants will be 
given by the elements in $R^{p+1}$ \cite{KarlMac}. If one wants to find the full line-element, the procedure can
therefore be
continued in the following way
 \begin{enumerate}
  \item[3.] Solve for the $\omega^{\alpha}$ (in a suitable numbering)
  and the $\omega^a_{\; b}$ from $R^{p+1}$.
  \item[4.] Determine the Lie algebra of the isometry group and some standard
  1-forms satisfying it.
  \item[5.] From these make an ansatz $\tilde {\omega}^i$ for the 1-forms over the full spacetime.
  \item[6.] Calculate the corresponding ${\tilde R}^{p+1}$ and compare 
  with $R^{p+1}$, giving differential or algebraic equations for the
  coefficients in the metric ansatz. These equations are of course equivalent 
  to (\ref{eq:Cartan4}). 
 \end{enumerate}

\section{Bianchi V and I}\label{sec:V}

In this section we consider homogeneous cosmological models of Bianchi type V and I, i.e.
those characterized by the symmetric matrix in the 
Ellis-MacCallum scheme \cite{EllisMacCallum} being zero. We assume 
that matter can be described as a perfect fluid. First the
preliminaries, like choice of frame and the elements in the set $S$ ($R^{p+1}$)
are given. Einstein's equations with a cosmological constant are used.
Then it is illustrated how one imposes 
the isometry group. After this we give the integrability conditions 
and reduce them to an integrable
system of five first order ordinary differential equations. 

\subsection{Preliminaries}

As energy-momentum tensor we take that of a perfect fluid
\begin{equation}\nonumber
T_{ij}=(\rho + p)u_iu_j-p\eta_{ij} 
\end{equation}
with linear equation of state $p=(\gamma - 1)\rho$. 
Here $\rho$ is the restframe density, $p$ is the isotropic pressure
and $u^i$ the 4-velocity of matter.
Since homogeneity
is assumed the elements in the set $S$ will depend on only one timelike
coordinate, that we choose as the density $\rho$. Sometimes, especially
in problems with more than one independent coordinate, it can be advantageous
to specify the coordinates at a later stage to simplify the equations, see \cite{Marklund}.

A Lorentz frame $\omega^i$ will be used. 
We choose a comoving frame, i.e., the 4-velocity is given by $u=\delta_i^0\omega^i=\omega^0$.
In general the normals of the hypersurfaces, $d\rho=\rho_{\vert i}\omega^i$, 
will be tilted relative to the 4-velocity. The 1-direction is chosen to be in the
direction of the spatial part of the density gradient, i.e., $\rho_{\vert 2}=\rho_{\vert 3}=0$.
From these equations we see that, once $\rho_{\vert 0}$ and $\rho_{\vert 1}$ are determined,
one can solve for $\omega^0$ as
\begin{displaymath}
\omega^0=\frac{d\rho}{\rho_{\vert 0}}-\frac{\rho_{\vert 1}}{\rho_{\vert 0}}\omega^1
\quad \hbox{or} \quad
X_0=\rho_{\vert 0}\frac{\partial }{\partial \rho} \, .
\end{displaymath}
This choice of frame means that we deviate from the usual approach of adopting the frame
to the hypersurfaces of homogeneity.

Since there is only one essential coordinate, $\rho$, this is the only 1-form that we will be able
to solve for from $R^{p+1}$.
The frame is finally fixed by requiring that the vorticity (rotation) vector of the fluid  is in the 12-plane, i.e., 
$\Omega^3\equiv\frac{1}{2}\epsilon^{3ijk}\omega_{ij}u_k=0$, corresponding
to that $\omega_{12}=0$ (see below for the definition of the vorticity tensor).  

The general model in this class will not have any isotropies and one can from the set 
consisting of $\rho_{|0}$, $\rho_{|1}$
and all $\gamma^i_{\,\, jk}$ construct the full set $R^{p+1}$. 
(The set could in principle be even more reduced since
the equations (\ref{eq:d2x}) give relations among the quantities.)
Since we want to impose Einstein's equations 
\begin{displaymath}
G_{ij}=T_{ij}+\Lambda\eta_{ij}
\end{displaymath}
for a perfect fluid and are using
a comoving frame the Einstein tensor 
should be given by 
\begin{displaymath}
G_{00}=\rho+\Lambda, \,\, G_{11}=G_{22}=G_{33}=p-\Lambda,
\end{displaymath}
where $\Lambda$ is the cosmological constant.
The cyclic identity must also be imposed. The nonzero elements
of the Riemann tensor are then given by
\begin{eqnarray}\nonumber
\vspace{1mm}
   &&R_{0101}=C_1-\frac{1}{2}p-\frac{1}{6}\rho+\frac{1}{3}\Lambda  , \, R_{0102}=R_{1323}=C_2  , \,\\ \nonumber
   &&R_{0103}=-R_{1223}=C_3  , \,  R_{0112}=R_{0323}=C_4  , \\ \nonumber
&&R_{0113}=-R_{0223}=C_5  , \,  R_{0123}=C_6  , \\ \nonumber   
&&R_{0202}=C_7-\frac{1}{2}p-\frac{1}{6}\rho+\frac{1}{3}\Lambda , \, R_{0203}=R_{1213}=C_8 , 
\\ \nonumber  
    &&R_{0212}=-R_{0313}=C_9  , \,  
R_{0213}=C_{10} , 
\\ \nonumber
  &&R_{0303}=-C_1-C_7-\frac{1}{2}p-\frac{1}{6}\rho+\frac{1}{3}\Lambda , \,  
    R_{0312}=C_{10}-C_6  ,  \\ \nonumber
 &&R_{1212}=C_1+C_7 -\frac{1}{3}\rho-\frac{1}{3}\Lambda , \,  R_{1313}=-C_7-\frac{1}{3}\rho
-\frac{1}{3}\Lambda , \\
    &&R_{2323}=-C_1-\frac{1}{3}\rho-\frac{1}{3}\Lambda  
\label{eq:Riemanncomp}
\end{eqnarray}
where $C_i$ are the ten independent components of the Weyl tensor.

Some of the rotation coefficients are expressible in terms of the kinematic
quantities shear $\sigma_{ij}$, vorticity $\omega_{ij}$, expansion $\theta$
and acceleration $a_i$
\begin{equation}\nonumber
  \gamma_{0ij} =
  -u_{i;j}=-\omega_{ij}-\sigma_{ij}+\frac{1}{3}h_{ij}\theta-a_{i}u_{j}. 
 \end{equation}
where 
$\sigma_{ij} = {h_i}^k{h_j}^l\left(u_{(k;l)} + \frac{1}{3}h_{kl}\theta \right)$,  
$\omega_{ij} = {h_i}^k{h_j}^l u_{[k;l]}$ , 
$\theta = u^i_{\,\, ;i}$ and $a_i=u_{i;j}u^j$ and $h_{ij}=u_iu_j-\eta_{ij}$
is the projection operator onto the space perpendicular to the 4-velocity.
In the next subsection we impose the requirement that the isometry group 
is of Bianchi type V or I. This will give six restrictions on the rotation coefficients.

\subsection{Symmetry group}

The Lie algebra of the isometry group can be determined by projection of Cartan's
equations onto the orbits \cite{KarlMac}. Since we do not have any isotropies in this case, the orbits
will be the hypersurfaces of homogeneity $d\rho=0$ in $M$. From $d\rho=0$ we get
\begin{displaymath}
d\rho = \rho_{\vert 0}\omega^0\vert+\rho_{\vert \alpha}\omega^{\alpha}\vert=0
\quad  \hbox{or}   \quad 
\omega^{0}\vert = -\frac{\rho_{\vert \alpha}}{\rho_{\vert 0}}\omega^{\alpha}\vert 
\end{displaymath}
where $\alpha=1,2,3$ are the spatial indices and a vertical bar $\vert$ indicates projection
onto $d\rho=0$. From the requirement of no isotropy  one has that
$\tau^i_{\,\, j}\vert =0$ holds on the
orbits and (\ref{connect}) then gives
\begin{displaymath}
\omega^i_{\,\, j}\vert= \gamma^i_{\,\, jk}\omega^k\vert \,\, .
\end{displaymath}
Hence the orbits are spanned by $\{\omega^1\vert,\omega^2\vert,\omega^3\vert\}$.
By projecting the first pair of Cartan's equations (\ref{Cartan1}) (the second pair will not give anything
new in this case) one obtains
\begin{equation}\nonumber
d\omega^{\alpha}\vert=\omega^j\vert\wedge\omega^{\alpha}_{\,\, j}\vert
\equiv \frac{1}{2}C^{\alpha}_{\,\, \beta \delta}\omega^{\beta}\vert\wedge\omega^{\delta}\vert
\end{equation}
where the structure constants are given by
\begin{equation}\label{eq:structureconstant}
\frac{1}{2}C^{\alpha}_{\,\, \beta \delta}=\gamma^{\alpha}_{\,\, [\beta \delta]}
+\frac{\rho_{\vert \beta}}{\rho_{\vert 0}}\gamma^{\alpha}_{\,\, [\delta 0]}
-\frac{\rho_{\vert \delta}}{\rho_{\vert 0}}\gamma^{\alpha}_{\,\, [\beta 0]} \, .
\end{equation}
The structure constants are hence given in terms of elements in $R^{p+1}$,
and since they are functions only of $\rho$ the $C^{\alpha}_{\,\, \beta \delta}$
are constants on the orbits.

From the Ellis-MacCallum scheme for the Bianchi classes, $\cite{RyanShep}$, we can decompose 
the structure constants into one symmetric $3\times 3$ matrix $N^{\alpha\beta}$
and a 3-vector $A^{\alpha}$ according to
\begin{eqnarray}\label{Nmatrix}\nonumber
\frac{1}{2}C^{\alpha}_{\,\, \beta \delta}\epsilon^{\beta\delta\gamma}=
N^{\alpha\gamma}+\epsilon^{\alpha\gamma\beta}A_{\beta}
\,\,\,\, \hbox{giving} \,\, \\ 
N^{\alpha\delta}=\frac{1}{2}\left[C^{\alpha}_{\,\, \beta \gamma}\epsilon^{\beta\gamma\delta} 
- C^{\gamma}_{\,\, \beta \gamma}\epsilon^{\beta\alpha\delta}\right] \, .
\end{eqnarray}
Bianchi classes V and I are characterized by $N^{\alpha\delta}=0$. This will give six relations
among the $\gamma^i_{jk}$. Note that we cannot simply use the standard form of the structure
constants in (\ref{eq:structureconstant}) and obtain nine relations for the Ricci coefficients this way
since our choice of frame (giving other relations among the elements in $S$) might be inconsistent with that giving the structure constants in standard
form. The nonzero Ricci rotation coefficients are then given by ($\alpha,\beta=1,2,3$ and $\omega_{12}=0$)
\begin{equation}\label{Riccirot}
\begin{array}{ll}
   \gamma_{0 \alpha 0} = - a_{\alpha}     , \,   
\gamma_{0 \alpha \beta} = - \sigma_{\alpha \beta} - \omega_{\alpha \beta}         \, \,\,
(\alpha \neq \beta), \\
\gamma_{0 \alpha \alpha} = \frac{1}{3} \theta - \sigma_{\alpha \alpha}      , \,      
  \gamma_{123} = \gamma_{132} = - \frac{\rho_{|1}}{\rho_{|0}} \sigma_{23}   , \\     
 \gamma_{133} = \gamma_{122} + \frac{\rho_{|1}}{\rho_{|0}} \left( \sigma_{11} + 2 \sigma_{22} \right)  , \,      
 \gamma_{231} = \frac{\rho_{|1}}{\rho_{|0}} \left( \gamma_{230} - \omega_{23} \right), \\  
  \gamma_{232} = \gamma_{131} + \frac{\rho_{|1}}{\rho_{|0}} \left( \sigma_{13} + \omega_{13} - \gamma_{130} \right)   , \,  \gamma_{130}  , \, \gamma_{131}       , \,  \gamma_{230},  \\   
\gamma_{233} = - \gamma_{121} + \frac{\rho_{|1}}{\rho_{|0}} \left( - \sigma_{12} + \gamma_{120} \right)  , \,        
 \gamma_{120} , \gamma_{121}  , \,  \gamma_{122}      .
\end{array}
\end{equation}

\subsection{Integrability conditions}

The commutator equations (\ref{eq:d2x}) give
\begin{equation}\label{eq:d2rho}
\frac{d\rho_{\vert i}}{d\rho}\rho_{\vert j}-\frac{d\rho_{\vert j}}{d\rho}\rho_{\vert i}
+\rho_{\vert k}(\gamma^k_{\,\, ji}-\gamma^k_{\,\, ij})=0
\end{equation}
and the Riemann equations (\ref{eq:Rie})
\begin{equation}
  {R^{i}}_{jkl} = 2{{\gamma}^{i}}_{j[l,\rho }{\rho}_{|k]} +
  2{{\gamma}^{im}}_{[l}{{\gamma}}_{jmk]} +
  2{{\gamma}^{i}}_{jm}{{\gamma}^{m}}_{[kl]}
\label{eq:Rie2}
 \end{equation} 
(antisymmetrization only over $kl$). The cyclic identity is already imposed due
to the choice (\ref{eq:Riemanncomp}) of the Riemann tensor and the Bianchi identities need not
be imposed due to the lack of isotropies. (However it is in practice often 
convenient to use the twice contracted Bianchi identities since they give simple
relations, but they may of course be derived from the other equations).
Hence the system (\ref{eq:d2rho}, \ref{eq:Rie2}) is the complete system. 
It is a set of first order ordinary differential equations, 
where the independent variable is $\rho$, and algebraic constraints. Some are
easily solved and give:
\begin{equation}\label{riccirot2}
\begin{array}{ll}
  a_2=a_3=\gamma_{121}=\omega_{23}=0    , \,   
   \gamma_{120}=\sigma_{12}   , \,      
    \gamma_{130}=\sigma_{13}+\omega_{13}  , \,     \\
 \gamma_{230}=-\sigma_{23} \,\,  (\hbox{or}\,\, \gamma_{131}=0)  , \,     
 {\rho_{\vert 0}}=-{\gamma\rho}\theta  , \,       
 a_1=(1-\gamma)v\theta  , \\  
 \omega_{13}=\frac{1}{2}v\gamma_{131}  \,      
\end{array}
\end{equation}
where we have introduced the tilt
\begin{equation}
v\equiv \frac{\rho_{\vert 1}}{\rho_{\vert 0}} \, .
\end{equation}
The case $\gamma_{131}=0$ will be treated separately in section \ref{irrotational}.
We see that the vorticity given by 
\begin{equation}\label{rotation}
\omega_{13}=\frac{1}{2}v\gamma_{131} 
\quad \hbox{or} \quad
\Omega_2 =-\frac{1}{2}v\gamma_{131} 
\end{equation}
is perpendicular to the acceleration, a result already found in \cite{KingEllis}.
The ten components, $C_i$, of the Weyl tensor are also given by (\ref{eq:Rie2}).

The system then reduces to nine first order differential equations
\begin{equation}\label{eq:diffeq}
\dot f_i(\rho)=F_i(f_j(\rho),\rho) \, ,
\end{equation}
where $\dot f \equiv df/d\rho$,
for the functions $f_i(\rho)$
\begin{equation}\nonumber
f_i \in \{v, \theta, \sigma_{11}, \sigma_{22}, \sigma_{12},
\sigma_{13}, \sigma_{23}, \gamma_{131}, B\} \, ,
\end{equation}
where we use the variable
\begin{equation}\label{Bdef}
B\equiv\gamma_{122}-
\frac{1}{3}v\theta + v \sigma_{22} 
\end{equation}
instead of $\gamma_{122}$,
and four algebraic constraints
\begin{equation}\label{eq:constraints}
G_a(f_j(\rho),\rho)=0 \, .
\end{equation} 
It turns out that the differentiated constraints, when using (\ref{eq:diffeq}), 
all are satisfied, i.e.,
\begin{displaymath}
\dot G_a = \frac{\partial G_a}{\partial f_j}\dot f_j + \frac{\partial G_a}{\partial \rho}=
 \frac{\partial G_a}{\partial f_j}F_j + \frac{\partial G_a}{\partial \rho}=0 \, .
\end{displaymath}
Hence the system can be reduced to five differential equations and 4 constraints
\cite{BradKarl}. 
The constraints are given by
\begin{eqnarray}\nonumber
&&\gamma_{131}\left[-2v^2\left(3\sigma_{11}+( 3\gamma -4)\theta\right)  
 - 3v B   - \right. \\ \nonumber
&&\left. 18  \left(\sigma_{11}+
\sigma_{22}\right)  \left( 1 -v^2\right)
\right] - 18 B  \sigma_{13} =0 , \\ \nonumber
&&\\ \nonumber
&&B \sigma_{12}  - \gamma_{131} \sigma_{23} \left(1- v^2\right)=0 , \\ \nonumber
&&\\ \nonumber
&& 9 \gamma_{131}  \left(\gamma_{131}v+2 \sigma_{13}  \right) \left( 1-v^2\right)+ 
6v\left(2 B^2 + \gamma\rho\right) 
 \\\nonumber
&&-2v^2B\left[2(3\gamma-4)\theta-3\sigma_{11}\right]-18B\sigma_{11}=0
\\ \nonumber
&& \\ \nonumber
&&9 \gamma_{131} \left[ \gamma_{131}\left(12-5v^2 \right)
-8v \sigma_{13}\right] +48v B \theta+\\ \nonumber
&& 36  \left[ \sigma_{11}^2+ \sigma_{11} \sigma_{22}+
\sigma_{22}^2+\sigma_{12}^2 +\sigma_{13}^2+\sigma_{23}^2+\Lambda+3B^2 \right. \\  \nonumber
&& \left.+\rho  -v^2\left(\sigma_{11} \sigma_{22}+\sigma_{22}^2+ \sigma_{23}^2\right)
\right]
+4\theta^2\left[(6\gamma-5)v^2-3\right] \\  
\label{constraints}
&&+12\sigma_{11} 
\left[6v B  +\left(3 \gamma -2\right) v^2\theta    \right]=0
\end{eqnarray}
that are easily solved for $\sigma_{13}$, $\sigma_{23}$ and $\sigma_{11}$ in which the three
first constraints are
linear. Finally the last constraint gives a second order polynomial in $\sigma_{22}$.
The differential equations are
\begin{eqnarray}\nonumber
\dot v &=& 
v\left(
  \frac{-\sigma_{11}}{\gamma\rho\theta} + \left(   \frac{4}{3\gamma} - 1 \right) \frac{1}{\rho} \right)	   \\ \nonumber
  \dot \gamma_{131} &=& \gamma_{131} \left(  \frac{ \left( \sigma_{11} + \sigma_{22} \right) }{\gamma\rho\theta} + \frac{1}{3\gamma\rho} \right)               \\ \nonumber
  \dot \sigma_{12} &=& \sigma_{12} \left( \frac{ \left( \sigma_{11} - \sigma_{22} \right) }{\gamma\rho\theta} + \frac{1}{\gamma\rho} \right) -
  	                  \frac{2 v\gamma_{131} \sigma_{23}}{\gamma\rho\theta}       \\ \nonumber
 \dot B &=& B\left(- \frac{\sigma_{11}}{ \gamma\rho\theta} + \frac{1}{3\gamma\rho}\right)+ 
\frac{2\gamma_{131}  \sigma_{13}} {\gamma\rho\theta} \\ \nonumber
(\theta^2)\dot &=& \left[3 \gamma_{131}^2 v^2 + 12 (\gamma-1) v\theta\left( B + v\sigma_{11}\right)-2\theta^2+ 
\right. \\ \nonumber
&&2(\gamma-1)(6\gamma-5)v^2 \theta^2-
3 \left( (3 \gamma-2) \rho-2\Lambda\right)  -
\\\nonumber
&& \left.
12 \left(\sigma_{11}^2  +
\sigma_{11} \sigma_{22}  + \sigma_{12}^2 + 
 \sigma_{13}^2 +  \sigma_{22}^2 +  \sigma_{23}^2 \right)\right] \\ \label{reducedsystem}
&&  /\left[3 \gamma \rho  \left((\gamma-1)
 v^2 - 1\right)\right]  
\end{eqnarray}

If one instead wants to use proper time for a comoving observer as independent coordinate
the equation
\begin{equation}
\frac{d\rho}{d\tau}=\rho_{\vert 0}=-\gamma\rho\theta \, ,
\end{equation} 
obtained by putting $dx=dy=dz=0$ in the line-element (see equation (\ref{eq:trialforms}) below), 
should be added to the system (\ref{reducedsystem}).
The derivatives in (\ref{reducedsystem}) are then expressed as
\begin{displaymath}
\dot f_i \equiv \frac{d f_i}{d \rho} = \frac{d f_i}{d \tau}\frac{d\tau}{d \rho} =
-\frac{d f_i}{d \tau}\frac{1}{\gamma\rho\theta} \ .
\end{displaymath}

The system has a unique solution for given initial conditions provided $F_i$ for the reduced
system (\ref{reducedsystem}) satisfy a Lipschitz condition (for example if $F_i$ are $C^1$
in a compact convex domain). The general solution hence depends on five constants of 
integration and in general its matter flow has both expansion, shear and vorticity.

One could of course choose to solve for four other quantities than 
$\sigma_{13}, \sigma_{23}, \sigma_{11}$ and $\sigma_{22}$ from (\ref{constraints}).
When doing a first order perturbative calculation around an isotropic universe, 
as in section \ref{sec:pert}, both $\sigma_{11}$ and $\sigma_{22}$ cannot be solved 
for from (\ref{constraints})
since the last constraint will be quadratic in $\sigma_{22}$ or $\sigma_{11}$.
Hence we solve for $\theta$ from the last constraint and use the differential equation
for $\sigma_{22}$ 
\begin{eqnarray}\nonumber
\dot \sigma_{22}&=& \frac{\dot\theta}{3} + 
\left[(\gamma - 2)\rho -4\left(\sigma_{12}^2-\left(v^2-1\right)\sigma_{23}^2\right)
-2\Lambda+
 \right. \\\nonumber
&&2\left(\sigma_{22}
-\frac{\theta}{3}\right)\left(v^2\left(2\sigma_{11}
+(6\gamma-5)\frac{\theta}{3}\right)-\theta\right)-4B^2  
\\\nonumber
&&+2v B \left(2(\sigma_{22}-\sigma_{11})-\gamma\theta\right)
+4\gamma_{131}\left(v\sigma_{13}-\gamma_{131}\right)\left.\right] \\\label{sigma22}
&&  /\left[2\gamma\rho\theta(v^2-1)\right]
\end{eqnarray}
instead of the one for $\theta$. 

This system is not in a suitable form for a dynamical system analysis. The system
should then be made autonomous and compact dimensionless variables should be introduced. 
In \cite{HewittWainwright} irrotational tilted Bianchi type V cosmologies were studied with this method.
The field equations are derived in terms of expansion-normalized variables making the state space compact. 
The existence of a monotonic function shows that the dynamics to a large extent is determined by the invariant 
subset of locally rotationally symmetric models. A complete analysis of the orbits with non-extreme tilt was obtained.
In \cite{ColeyHervik} tilted Bianchi models of solvable type, including Bianchi type V, were considered with emphasis on the late-time behaviour. 
The equilibrium points were given in \cite{HewittWainwright}, but here the stability analysis was performed in the full state space. 
It was found that for $2/3<\gamma<2$ Bianchi type V models approach the Milne universe in the asymptotic future.
To complete the analysis a study of the behaviour for early times (high densities) would be of interest.

\subsection{The metric}

From equations  (\ref{eq:structureconstant}) for the structure constants we find the Lie-algebra of the isometry group to be
\begin{eqnarray}\nonumber
    d\omega^1\vert    &   =     &  \gamma_{131}\omega^1\vert \wedge \omega^3\vert        \\ \nonumber
    d\omega^2\vert    &   =     &   B\omega^1\vert \wedge \omega^2\vert +
 \gamma_{131}\omega^2\vert \wedge \omega^3\vert           \\
    d\omega^3\vert    &   =     &   B \omega^1\vert \wedge \omega^3\vert         
\end{eqnarray}
where $B$ is given by (\ref{Bdef}).
As expected it is not in the standard form for Bianchi V, unless $B=0$. Guided by the
above algebra and standard 1-forms for Bianchi V, we make the following simplified, but 
sufficient, ansatz for a basis of 1-forms
for the full spacetime
\begin{eqnarray}\nonumber
 \tilde \omega^0       &=&  -\frac{d\rho}{\gamma\rho\theta}-v\tilde\omega^1 
\,\, , \quad    \tilde \omega^1   =  f_1 e^{-x} dy 
\\ \nonumber
   \tilde \omega^2   &  =  & g_1 e^{-x} dy + g_2 e^{-x} dz + g_3 dx  
\\
 \tilde \omega^3   & = &  \frac{1}{\gamma_{131}} dx + h_1\frac{B}{\gamma_{131}} e^{-x} dy 
\label{eq:trialforms}
\end{eqnarray}
where $f_1$, $g_1$, $g_2$, $g_3$ and $h_1$ are functions of $\rho$ to be determined.
(The case $\gamma_{131}=0$ is treated separately in section \ref{irrotational}.)
From (\ref{eq:trialforms}) we calculate the set $\tilde S =\{\tilde \rho_{\vert i}, \, \tilde 
\gamma_{ijk}\}$. A comparison with the set $S$ gives
\begin{eqnarray}\nonumber
  f_1 &=& h_1 = C_1  \rho^{\frac{1}{\gamma}-1}v \quad , \quad  \,\, 
 g_1 = \gamma_{131}\rho^{1- \frac{2}{\gamma}}v I_1 
      \\ 
g_2 &=& C_2 \gamma_{131}\rho^{1- \frac{2}{\gamma}}v
  \,\, , \quad
g_3 = \gamma_{131}\rho^{1- \frac{2}{\gamma}}v I_2
\label{eq:difftrial}
\end{eqnarray}
where $C_1$ and $C_2$ are (nonzero) constants of integration that can be absorbed in the
definitions of the coordinates $y$ and $z$ and the integrals $I_1$ and $I_2$
are given by
\begin{eqnarray}\nonumber
I_1 &=& \frac{2C_1}{\gamma}\int \frac{\rho^{\frac{3}{\gamma}-2}}{v^2\rho\theta\gamma^2_{131}}
\left[\sigma_{12}\gamma_{131}+
\sigma_{23}B
\right]d\rho \\ \label{eq:integral}
I_2 &=& \frac{2}{\gamma}\int\frac{\sigma_{23}\rho^{\frac{2}{\gamma}-1}}{v\rho\theta\gamma^2_{131}}d\rho
\end{eqnarray}
respectively. From the 1-forms the metric is given by
$ds^2=\eta_{ij}\omega^i \omega^j$.
The full metric is hence given in terms of quadratures once the set $S$ has been
constructed.
The solution depends on some arbitrary constants of integration and an even more general ansatz than (\ref{eq:trialforms}) could have been made, but all these metrics give the same set $S$ and
hence they are locally equivalent. 

Later on we will consider perturbations around the open Friedmann model, and will be interested 
in the limit $v \rightarrow 0$ as $\epsilon \rightarrow 0$. As seen some of the components
in the metric will then diverge. This coordinate singularity can be avoided by changing coordinates
according to
\begin{displaymath}
y=\epsilon \tilde y \,\, , \quad z=\frac{\tilde z}{\epsilon} \,  .
\end{displaymath}

\subsection{Orthogonal solutions}

For the orthogonal case $v=0$, when also the vorticity is zero, 
it is better to use a frame where the shear tensor
is diagonal, $\sigma_{12}=\sigma_{13}=\sigma_{23}=0$. 
The obtained system is easily solved and all solutions are well known.
Essentially only two types of solutions appear. These are the Bianchi V solutions
\begin{eqnarray}\nonumber
   \sigma_{11}&=&-\sigma_{22}=-k_1\rho^{\frac{1}{\gamma}}\, ,\;\;
   \gamma_{131}=\gamma_{232}=k_2\rho^{\frac{1}{3\gamma}}\;,\\\label{orthoV}
   \theta&=&\pm\sqrt{3}\sqrt{3k_2^2\rho^{\frac{2}{3\gamma}}+k_1^2\rho^{\frac{2}{\gamma}}
+\rho+\Lambda}
\end{eqnarray}
and the Bianchi class I solutions
\begin{eqnarray}\nonumber
   \sigma_{11}&=& k_1\rho^{\frac{1}{\gamma}}\, , \;\;   \sigma_{22}=k_2\rho^{\frac{1}{\gamma}}\\  
   \theta&=&\pm\sqrt{3}\sqrt{(k_1^2+k_2^2+k_1k_2)\rho^{\frac{2}{\gamma}}+\rho+\Lambda} \, .
\end{eqnarray}
All orthogonal solutions can be found from these interchanging the 1-, 2- and 3-directions.\\ 
The corresponding metrics are
\begin{eqnarray}\nonumber
   \omega^0&=&-\frac{d\rho}{\gamma\rho\theta},\;\; \omega^1=c_1e^{-\int{\frac{\gamma_{011}}{\gamma\rho\theta}}d\rho}e^{-x}dy, \\ \nonumber 
   \omega^2&=&c_2e^{-\int{\frac{\gamma_{022}}{\gamma\rho\theta}}d\rho}e^{-x}dz,\;\; 
\omega^3=\frac{1}{k_2}\rho^{-\frac{1}{3\gamma}}dx
\end{eqnarray}
and
\begin{eqnarray}\nonumber
   \omega^0&=&-\frac{d\rho}{\gamma\rho\theta} \;,\;\; \omega^1=c_1e^{-\int{\frac{\gamma_{011}}{\gamma\rho\theta}}d\rho}dx \;,\\ \nonumber   
   \omega^2&=&c_2e^{-\int{\frac{\gamma_{022}}{\gamma\rho\theta}}d\rho}dy \;,\;\; 
\omega^3=c_3e^{-\int{\frac{\gamma_{033}}{\gamma\rho\theta}}d\rho}dz
\end{eqnarray}
respectively.
The integrals can be performed for specific values of $\gamma$.

\subsection{Irrotational tilted solutions}\label{irrotational}

In \cite{HewittWainwright} this class for non-extreme tilt ($v<1$) was studied using a dynamical systems approach. 
It was found that that the dynamics to a large extent is determined by the invariant subset of locally rotationally
symetric (LRS) models.
This subset was studied in detail in \cite{CollinsEllis}  for different values of $\gamma$. 
There are solutions for which the tilt always is less than one and hence 
the hypersurfaces of homogeneity remain spacelike when going backwards in time. 
However, it may approach one for large times for certain values of $\gamma$ including the
radiation case $\gamma=4/3$.
There are also solutions for which the tilt gets larger than one for small times and hence the hypersurfaces change from 
being spacelike to being timelike. For this class singularities may occur for finite densities. For example, 
in the $\gamma=4/3$ case some of the kinematic quantities and the Weyl tensor diverge for $v=\sqrt{3}$.

The irrotational solutions are given by $\gamma_{131}=0$, which implies that the vorticity vanishes and the frame cannot be fixed by demanding $\Omega^3=0$. Instead the frame is fixed by putting
$\sigma_{23}=0$. Some care should be taken in using equations (\ref{constraints}) and
(\ref{reducedsystem}) directly since an extra constraint is introduced and they were derived
assuming $\gamma_{131}\neq 0$. Yet the only nontrivial cases are those obtained from this
system with $\gamma_{131}=\sigma_{23}=0$ and $B\neq 0$. From the two first constraints
$\sigma_{12}=\sigma_{13}=0$ is obtained. $\sigma_{11}$ and $\sigma_{22}$ can be solved
for from the two others. The system of differential equations is reduced to a system for
$\dot v$, $\dot B$ and $\dot\theta$. This can be reduced to a system of two
differential equations since the equations for  $\dot v$
and $\dot B$ in this case can be combined to 
\begin{displaymath}
\frac{\dot  v}{ v} =
\frac{ \dot B}{B}
   + \left( \frac{1}{\gamma} - 1 \right) \frac{1}{\rho} 
\end{displaymath}
with solution
\begin{equation}\label{Bbeta}
B =C_1  v \rho^{\frac{\gamma-1}{\gamma}}
\end{equation}
where $C_1$ is a constant of integration.
The metric is in this case given by:
\begin{eqnarray}
	 \nonumber \omega^0&=&-\frac{d\rho}{\gamma\rho\theta}- v\omega^1,
	 \quad \omega^1=-\frac{1}{B}dx\\ \nonumber
	 \nonumber \omega^2&=&k_2\rho^{-\frac{1}{3\gamma}}e^{\int\frac{\sigma_{22}}{\gamma\rho\theta}d\rho}e^{-x}dy, 
\\ \nonumber
\omega^3&=&k_3\rho^{-\frac{1}{3\gamma}}e^{-\int\frac{\sigma_{11}+\sigma_{22}}{\gamma\rho\theta}d\rho}e^{-x}dz \, .
\end{eqnarray}

\subsubsection{LRS solutions}

LRS solutions are obtained by putting $\sigma_{11}=-2\sigma_{22}$ in the above equations
(the LRS symmetry lies in the 23-plane). If one solves the constraints for $\sigma_{22}$ and $\theta$ the system is reduced to one differential
equation for $ v$.

For $\gamma=1$ and $\Lambda=0$
the equations can be completely integrated
\cite{Farnsworth,MarkBrad}. 
The solution in our variables is given by
\begin{eqnarray} \nonumber
\sigma_{11}&=&-2\sigma_{22}=
\pm\frac{\kappa^2\left(3C\kappa+2\right)}{3\left(D-\kappa\sqrt{1+C\kappa}\right)} \;,\\ \nonumber
\nonumber
\theta&=&\mp\frac{3D\kappa\sqrt{1+C\kappa}-\kappa^2\left(2+\frac{3}{2}C\kappa\right)}{D-\kappa\sqrt{1+C\kappa}} \\   
 v&=&\mp\frac{\kappa}{D-\kappa\sqrt{1+C\kappa}}, \;\;  \gamma_{122}=-\kappa
\end{eqnarray}
where $C$ and $D$ are constants of integration and $\kappa$ and $\rho$ are related as
\begin{eqnarray}\label{densdust}
   \rho=\frac{3CD\kappa^3}{D-\kappa\sqrt{1+C\kappa}} \, .
\end{eqnarray}
The basis 1-forms are
\begin{eqnarray} \nonumber
\omega^0&=&\pm\frac{d\kappa}{\kappa^2\sqrt{1+C\kappa}}\mp dx
\;, \;
\omega^1=-\frac{D-\kappa\sqrt{1+C\kappa}}{\kappa}dx
\;,\;  \\ \omega^2&=&\frac{e^{Dx}}{\kappa}dy \;,\;\;
   \omega^3=\frac{e^{Dx}}{\kappa}dz \, .
\end{eqnarray}

From (\ref{densdust}) one finds essentially two types of solutions. If both $C$ and $D$ are positive the density
rises from zero to infinity when $\kappa$ goes from zero to $\kappa_1$, where $\kappa_1$ is given by
\begin{displaymath}
D-\kappa_1\sqrt{1+C\kappa_1}=0 \, .
\end{displaymath}
The tilt $ v$ then also increases from zero to infinity, and hence the hypersurfaces of homogeneity
change from being spacelike to being timelike. 

A positive density is also obtained for $C>0$ and $D<0$, and in this
case the density goes from zero to infinity when $\kappa$ goes from zero to infinity, but now the tilt
varies from zero to zero in this interval. The maximum value is obtained for $\kappa_2$ 
\begin{displaymath}
C\kappa_2^2+2D\sqrt{1+C\kappa_2}=0 \, .
\end{displaymath}
giving 
\begin{displaymath}
\vert v_{max}\vert = \frac{\sqrt{1+C\kappa_2}}{1+\frac{3}{2}C\kappa_2} 
\end{displaymath}
that is less then one and hence the hypersurfaces of homogeneity remain spacelike for all times.

A perturbative calculation (see \ref{sec:pert} for the perturbative method) with a small 
$
\gamma_{131}=\epsilon\gamma^{ \left( 1 \right) }_{131} \, 
$
around the exact solution gives
\begin{equation}
\gamma^{ \left( 1 \right) }_{131} = F \frac{\kappa}{D-\kappa\sqrt{1+C\kappa}}
\end{equation}
where $F$ is a constant of integration. Hence the perturbation grows as the other quantities
for the first case ($C,D>0$) as $\kappa \rightarrow \kappa_1$. However, for the second case ($D<0$) it
goes to zero when $\kappa \rightarrow \infty$.

\subsection{Solutions with a timelike homothetic motion}

A spacetime has a homothetic Killing vector $\xi$ if
\begin{equation}\nonumber
\xi_{\mu;\nu}+ \xi_{\nu;\mu}=2cg_{\mu\nu}
\end{equation}
is satisfied for some constant $c$. This implies that quantities of the same dimension scale
in the same way in the direction of $\xi$. For a timelike homothetic Killing vector it is hence
easy to show that for perfect fluids all components of the Riemann tensor, $R_{ijkl}$, are proportional to
$\rho$, the Ricci rotation coefficients to $\rho^{1/2}$ and $\rho_{|i}$ to $\rho^{3/2}$, see
e.g. \cite{RosquistJantzen}. The field equations are then reduced to algebraic equations. 

Unfortunately this limitation is quite restrictive and the only solution of this type is the
flat Friedmann universe (of Bianchi type I).

\section{Perturbative solutions}\label{sec:pert}

In this section we consider perturbative solutions to the system (\ref{constraints}) and (\ref{reducedsystem}).
In general perturbative solutions to systems of non-linear equations with constraints need not correspond to 
exact solutions, but in the present case the system can be reduced to (\ref{reducedsystem}), that under reasonable
assumptions is known to be integrable with five constants of integration. Hence a perturbative solution, obtained
by solving the Taylor expanded equations, should agree with the Taylor expansion of some exact solution. This is
also favoured by comparing the perturbative solutions with a numerical solving of the full system, 
see section \ref{sec:numerical}.

We here first briefly recall some basic results of perturbation theory.
Calling the four functions solved for from the algebraic constraints (\ref{eq:constraints}) $g_a$, $a,b,..=1,...,4$,
the reduced system can be written as
\begin{eqnarray}\nonumber
\dot f_i&=&F_i(f_j,g_a,\rho), \,\; i,j,..=1,...,5 \, ,\\
\label{redsys2}
0&=&G_a(f_j,g_b,\rho) \, .
\end{eqnarray} 
The functions $f_i$ and $g_a$ are expanded around the solution $f^{(0)}_i$ and $g^{(0)}_a$ of (\ref{redsys2})
in the small parameter $\epsilon$ as
\begin{eqnarray}\nonumber
\Delta f_i \equiv f_i - f^{(0)}_i = \epsilon f^{(1)}_i +\epsilon^2 f^{(2)}_i+... \\\label{expansion1}
\Delta g_a \equiv g_a - g^{(0)}_a = \epsilon g^{(1)}_a +\epsilon^2 g^{(2)}_a+...
\end{eqnarray}
giving, with $H_p$ equal to $F_i$ or $G_a$, 
\begin{eqnarray}\nonumber
H_p &=& H_p(f^{(0)}_j,g^{(0)}_a,\rho) + \frac{\partial H_p}{\partial f_j}\Delta f_j +
\frac{\partial H_p}{\partial g_a}\Delta g_a +\\\nonumber
&&\frac{\partial^2 H_p}{\partial f_j \partial g_a}\Delta f_j \Delta g_a 
+\frac{1}{2}\frac{\partial^2 H_p}{\partial f_j \partial f_k}\Delta f_j \Delta f_k + \\\nonumber
&&\frac{1}{2}\frac{\partial^2 H_p}{\partial g_a \partial g_b}\Delta g_a\Delta g_b+...
=H_p(f^{(0)}_j,g^{(0)}_a,\rho) +\\\nonumber
&& \epsilon \left(\frac{\partial H_p}{\partial f_j}f^{(1)}_j +\frac{\partial H_p}{\partial g_a} g^{(1)}_a \right)+
\epsilon^2\left(\frac{\partial H_p}{\partial f_j}f^{(2)}_j +\right.\\\nonumber
&&\left.\frac{\partial H_p}{\partial g_a} g^{(2)}_a \right)
+\epsilon^2\left(\frac{1}{2}\frac{\partial^2 H_p}{\partial f_j \partial f_k}f^{(1)}_j f^{(1)}_k+ \right.\\\label{expansion2}  
&&\left.\frac{1}{2}\frac{\partial^2 H_p}{\partial g_a \partial g_b} g^{(1)}_ a g^{(1)}_ b+ 
 \frac{\partial^2 H_p}{\partial f_j \partial g_a}f^{(1)}_j g^{(1)}_ a
 \right)+...
\end{eqnarray}
where the partial derivatives are evaluated at $f_i = f^{(0)}_i $ and $g_a = g^{(0)}_a$. 
Identifying equal powers of $\epsilon$ in (\ref{redsys2}), using (\ref{expansion1}) and
(\ref{expansion2}) with $H_p=G_a$, one can solve for $g^{(n)}_a$ in terms of $f^{(n)}_i$ and lower order quantities as
\begin{equation}\label{g}
g^{(n)}_a =-\left({G^{-1}}\right)^b_a\frac{\partial G_b}{\partial f_j} f^{(n)}_j+{\cal{G}}_a^{(n-1)}
\end{equation}
where $\left({G^{-1}}\right)^b_a$ is the inverse of $\frac{\partial G_a}{\partial g_b}$ (assuming that it is invertable) and ${\cal{G}}_a^{(n-1)}$ 
only depends on $f_i^{(m)}$ and $g_a^{(m)}$ up to order $n-1$, i.e. $m\leq n-1$. Substitution of (\ref{g}), (\ref{expansion1})
and (\ref{expansion2}) with $H_p=F_i$ into (\ref{redsys2}) and identification of equal powers of $\epsilon$ now gives
\begin{equation}\label{norderdiff}
\dot f_i^{(n)}-\left(\frac{\partial F_i}{\partial f_j}-\frac{\partial F_i}{\partial g_a}\left({G^{-1}}\right)^b_a \frac{\partial G_b}{\partial f_j}\right)f_j^{(n)}=
{\cal{F}}_i^{(n-1)}
\end{equation}
where ${\cal{F}}_i^{(n-1)}$ only depends on $f_i^{(m)}$ and $g_a^{(m)}$ up to order $n-1$ and for 
$n=1$ reduces to ${\cal{F}}_i^{(n-1)}={\cal{F}}_i^{(0)}=0$.
From this equation we recall the result that the homogeneous parts of the differential equations are the same to each order
in $\epsilon$, and hence the integration constants add up as $k_i=\epsilon k_i^{(1)}+\epsilon^2 k_i^{(2)}+...$, so that the correct
number of constants is maintained to any order.

\subsection{First order perturbations}

We here consider perturbations to first order  in the small parameter $\epsilon$ around
the open Friedmann universe. The
nonzero elements in the set $S$  for the Friedmann universe are given by
\begin{eqnarray}\label{eq:0th}\nonumber
\theta^{\left( 0 \right)} &=& \pm \sqrt{3} \sqrt{ 3 \left( k_{1}^{ \left( 0 \right) } \right)^{2} \rho^{ \frac{2}{3 \gamma} } + \rho +\Lambda},\\ \nonumber
\gamma^{\left( 0 \right)}_{133}&=&\gamma^{\left( 0 \right)}_{122}, \\
\gamma^{\left( 0 \right)}_{131} &=& \gamma^{\left( 0 \right)}_{232} =\pm \sqrt{  \left( k_{1}^{ \left( 0 \right) } \right)^{2} \rho^{ \frac{2}{3 \gamma} } - (\gamma^{\left( 0 \right)}_{122})^2 }
\end{eqnarray}
where $k_{1}^{\left( 0 \right)}$ is a constant of integration. In the following we choose 
$\gamma^{\left( 0 \right)}_{122}=0$, so that 
$\gamma^{\left( 0 \right)}_{131} = \gamma^{\left( 0 \right)}_{232}= k_{1}^{\left( 0 \right)} \rho^{ \frac{1}{ 3 \gamma } }$. The freedom in one of the Ricci rotation coefficients is due to that only the 
$\gamma_{0ij}$ appear in $S$ for the Friedmann universe due to the isotropy.
Note, however, that the resulting perturbed solutions are depending on this choice.
For example, if we instead had chosen $\gamma^{\left( 0 \right)}_{131}=0$, the
first order perturbations would have had zero vorticity.

Instead of solving for $\sigma_{22}$ from the last of equations (\ref{constraints})
we solve this equation for $\theta$, and hence the differential equation for $\theta$
in (\ref{reducedsystem}) is replaced by the corresponding for $\sigma_{22}$, (\ref{sigma22}). 

To first order we write the elements in $S$ as
\begin{equation}\nonumber
 v=\epsilon  v^{ \left( 1 \right) }\, , \,\,
\gamma_{ijk}=\gamma^{ \left( 0 \right) }_{ijk}+\epsilon\gamma^{ \left( 1 \right) }_{ijk} \, .
\end{equation}
Expanding the system (\ref{constraints}) and (\ref{reducedsystem}) (with the last equation replaced by 
(\ref{sigma22}))
to first order then gives  
\begin{eqnarray}\nonumber
 a_1& = & \epsilon k_{3} \theta \rho^{ \frac{4}{3 \gamma} -1 } \left( 1 - \gamma \right) , \\ \nonumber
\sigma_{11}&=&-\sigma_{22}=\epsilon k_{2} \rho^{ \frac{1}{\gamma} }, \,\,\, 
 \sigma_{12}= \epsilon k_{4} \rho^{ \frac{1}{\gamma} }, \\ \nonumber
\sigma_{13} &=& -\epsilon\frac{\gamma k_{3}}{ 3  k_{1} }   \rho^{ \frac{1}{\gamma} } - \epsilon\frac{k_1 k_{3}}{2  }   \rho^{ \frac{5}{3 \gamma} - 1 },
\\ \nonumber
\omega_{13} &=& \epsilon \frac{k_1 k_{3}}{2  }   \rho^{ \frac{5}{3 \gamma} - 1 },\;\;
 \gamma_{131} =  k_{1} \rho^{ \frac{1}{ 3 \gamma } },
\end{eqnarray}
\begin{eqnarray}\nonumber
\gamma_{122} &=&  \epsilon \gamma_{122}^{ \left( 1 \right) }\quad \hbox{where} \quad
\gamma^{(1)}_{122} = k_{5} \rho^{ \frac{1}{3 \gamma} } - \\ \nonumber
&&\frac{  k_{3} }{6 } \rho^{ \frac{1}{3 \gamma} } 
\int{\left[\frac{2}{\gamma}(\gamma-1)\theta\rho^{\frac{1}{\gamma}-2}
+\frac{\rho^{\frac{1}{\gamma}-1}}{\theta}\right] d\rho } \\\nonumber
  \theta &=& \pm \sqrt{3}  \sqrt{ 3 k_{1}^{2} \rho^{ \frac{2}{3 \gamma} } + \rho+\Lambda }, \\ 
   v &=& \epsilon  k_{3} \rho^{ \frac{4}{3 \gamma} - 1 } 
\label{eq:Sperturb}
\end{eqnarray}
where $k_1=k_{1}^{\left( 0 \right)}+\epsilon k_{1}^{\left( 1 \right)}$, $k_{2}=k_{2}^{\left( 1 \right)}$, 
$k_{3}=k_{3}^{\left( 1 \right)}$, $k_{4}=k_{4}^{\left( 1 \right)}$ and $k_{5}=k_{5}^{\left( 1 \right)}$
are the five constants of integration. 

The integral in $\gamma_{122}$ can be evaluated for certain values of $\gamma$
if $\Lambda=0$. For $\gamma=1$
(dust) one gets
\begin{eqnarray}\nonumber
  \gamma_{122}^{ \left( 1 \right) } = k_{5} \rho^{ \frac{1}{3} } \pm \frac{ k_{3} \theta }{ 9  } \left( \rho^{ \frac{1}{3} } - 6 k_{1}^{2} \right) \, ,
\end{eqnarray}
for $\gamma=\frac{4}{3}$ (radiation)
\begin{eqnarray}\nonumber
&&  \gamma_{122}^{ \left( 1 \right) } = k_{5} \rho^{ \frac{1}{4} } -
\\ \nonumber
&&k_{3} \rho^{ \frac{1}{4} }  \left[ 
 \frac{5}{9}\theta\rho^{-\frac{1}{4}} 
\pm 
k_{1}\ln \left( \frac{  \rho^{ \frac{1}{4} }\pm\frac{1}{\sqrt{3}}\theta\rho^{-\frac{1}{4}}
 - \sqrt{3} k_{1} }
{\rho^{ \frac{1}{4} }\pm\frac{1}{\sqrt{3}}\theta\rho^{-\frac{1}{4}} + \sqrt{3} k_{1} } \right)
\right]
\end{eqnarray}
and for $\gamma=2$ (stiff matter), with $\tilde k_{5}=k_{5} \pm k_{3}/2 $,
\begin{eqnarray}\nonumber
&&  \gamma_{122}^{ \left( 1 \right) } = \tilde k_{5} \rho^{ \frac{1}{6} } \mp 
\\ \nonumber
&&\frac{\sqrt{3} k_{3} \rho^{ \frac{1}{6} } }{6} 
\left[ 4 \ln \left(\frac{\rho^{\frac{1}{3}}}{\sqrt{3}k_1}\pm\frac{\rho^{-\frac{1}{6}}\theta}{3k_1}\right)
-\frac{9\sqrt{3}k_1^2\rho^{-\frac{1}{6}}}{\sqrt{3}\rho^{\frac{1}{2}}\pm\theta}\right] \ .
\end{eqnarray}
The integral can also be evaluated with nonzero $\Lambda$ for $\gamma=1$ if $k_1=0$, corresponding to
that the background metric is of Bianchi type I (but the perturbed one is of type V), giving
\begin{displaymath}
 \gamma_{122}^{ \left( 1 \right) } = k_{5} \rho^{ \frac{1}{3} }\mp \frac{k_3}{3\sqrt{3}}\rho^{\frac{1}{3}}\sqrt{\rho+\Lambda} \, .
\end{displaymath}

\subsection{The metric to first order}

The form of the metric (\ref{eq:trialforms}-\ref{eq:integral}) is not suitable for a perturbative calculation since some
of the coefficients diverge when $\epsilon \rightarrow 0$. This can be avoided by introducing
new coordinates according to
\begin{equation}
y=\epsilon \tilde y \,\, , \quad z=\frac{ \tilde z}{\epsilon} \ .
\end{equation}
When expanding the 1-forms to first order in $\epsilon$ the tilt $ v$ to second order, given in \ref{section:2nd},
will be needed. 
To first order in $\epsilon$ one then obtains the following 1-forms (with $C_1=C_2=1$)
\begin{eqnarray}\nonumber
\omega^0 &=&- \frac{d\rho}{\gamma\rho\theta}-\epsilon\rho^{\frac{1}{\gamma}-1}e^{-x}d\tilde y \\ \nonumber
\omega^1&=&\frac{1}{k_{3}}\rho^{-\frac{1}{3\gamma}}e^{-x}\left(1-\epsilon A_1\right)
d\tilde y \\ \nonumber
\omega^2&=& -\epsilon\frac{2  k_{4}}{k_{2} k_{3}}A_1\rho^{-\frac{1}{3\gamma}}e^{-x}d\tilde y+ \\ \nonumber
&&k_1 k_{3}\rho^{-\frac{1}{3\gamma}}e^{-x}\left(1+\epsilon A_1\right)d\tilde z 
\\ \nonumber
\omega^3&=&\frac{1}{k_1}\rho^{-\frac{1}{3\gamma}}dx
+\\ \label{formsperturb}
&&\frac{\epsilon}{k_1}e^{-x}
\left(\frac{1}{k_{3}}\rho^{-\frac{2}{3\gamma}}\gamma_{122}^{(1)}- 
\frac{1}{3}\rho^{\frac{2}{3\gamma}-1}\theta\right)d\tilde y
\end{eqnarray}
where $\theta$ and $\gamma_{122}^{(1)}$ are given by (\ref{eq:Sperturb}) and $A_1$ by
\begin{equation}
A_1(\rho)=-k_{2} \int \frac{\rho^{\frac{1}{\gamma}-1}}{\gamma\theta}d\rho  \,\, \  .
\end{equation}
For dust ($\gamma=1$) and $\Lambda=0$ $A_1$ becomes
\begin{equation}\nonumber
A_1= -\frac{2 k_{2} \theta}{3 }
\left(1-6 k_1^2\rho^{-\frac{1}{3}} \right) 
\end{equation}
and with $\Lambda \neq 0$ and $k_1=0$
\begin{equation}\nonumber
A_1= -\frac{2 k_{2} \theta}{3 } \, .
\end{equation}
Radiation ($\gamma=4/3$) with $\Lambda = 0$ gives
\begin{equation}\nonumber
A_1= - k_{2}\theta \rho^{-1/2} \, .
\end{equation}

The relation to proper time for a comoving observer is obtained by putting $dx=dy=dz=0$
in the line-element and is given by
$d\tau=-\frac{d\rho}{\gamma\rho\theta}$. 
For example, for an expanding universe with $\gamma=1$ we have 
$\theta=\sqrt{3} \sqrt{ 3 k_{1}^{2} \rho^{ \frac{2}{3} } + \rho }$. 
Integration gives (with integration constant such that $\tau(\rho = \infty)=0$)
\begin{displaymath}
\tau = \frac{\theta \rho^{-\frac{2}{3}}}{3 k_1^2}-
\frac{1}{6 k_1^3}\ln\left(\frac{\theta + 3 k_1 \rho^{\frac{1}{3}}}
{\theta - 3 k_1 \rho^{\frac{1}{3}}} \right) \, .
\end{displaymath}
Note that deviation from zeroth order quantities in the expression for proper time will appear first to second order.

\subsection{Second order perturbations}\label{section:2nd}

From (\ref{g}) we see that the second order perturbations can be obtained up to quadratures.
A full second order calculation will be done in a forthcoming paper. Here we focus on the tilt, $ v$,
whose $n$:th order equations decouple from other $n$:th order quantities, and also since it will be needed
to second order to obtain the metric to first order.

To second order one obtains
\begin{equation}
 v = \epsilon k_3 \rho^{\frac{4}{3\gamma}-1}
\left(1-\frac{\epsilon k_2}{\gamma}\int \frac{\rho^{\frac{1}{\gamma}-1}}{\theta}d\rho\right) 
\end{equation}
where now $k_3=k^{(1)}_3+\epsilon k^{(2)}_3$. For $\Lambda=0$ and $\gamma=1$ 
we have
\begin{eqnarray}\nonumber
 v &=& \epsilon k_3 \rho^{\frac{1}{3}}
\left(1-\frac{2\epsilon k_2}{3}\theta
(1-6 k_1^2\rho^{-\frac{1}{3}})\right) =\\ \nonumber
&&\epsilon k_3 \rho^{\frac{1}{3}}
\left(1\mp\frac{2\epsilon k_2}{\sqrt{3}}
\sqrt{\rho^{\frac{1}{3}}+3k_1^2}
(\rho^{\frac{1}{3}}-6 k_1^2)
\right) 
\end{eqnarray}
and for $\Lambda\neq 0$, $\gamma=1$ and $k_1=0$
\begin{eqnarray}\nonumber
 v &=& \epsilon k_3 \rho^{\frac{1}{3}}
\left(1\mp\frac{2\epsilon k_2}{\sqrt{3}}
\sqrt{\rho+\Lambda}
\right) \, .
\end{eqnarray}
For $\Lambda=0$ and $\gamma=4/3$ the tilt is given by
\begin{eqnarray}\nonumber
 v &=& \epsilon k_3 
\left(1-\epsilon k_2\theta\rho^{-\frac{1}{4}}\right)= \\ \nonumber
&&\epsilon k_3 \left(1\mp\epsilon k_2\sqrt{3}
\sqrt{\rho^{\frac{1}{2}}+3k_1^2}
\right) \, .
\end{eqnarray}
These results are in accordance with the exact equation (\ref{reducedsystem}) for $ v$  that
shows that the behaviour is determined by the sign of $\sigma_{11}/\theta$ 
($\sigma_{11}=\epsilon k_2\rho^{1/\gamma}$).

\section{Numerical solutions}\label{sec:numerical}

In this section we solve the system (\ref{constraints}-\ref{reducedsystem}) numerically using the Runge-Kutta method with a truncation
error of order $h^4$ in the step length $h$. 
A comparison between the code and the exact solutions (\ref{orthoV}) gives agreement to high accuracy.

\subsection{Comparison with perturbative calculations}

\begin{figure}[h]
\epsfxsize=3in
\epsffile{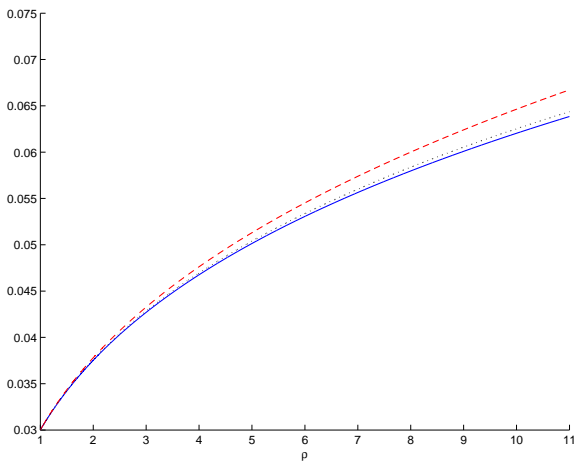}
\vskip3mm
\caption{Comparison between numerical calculation (solid curve) and perturbations to first (dashed) and second order 
(dotted) for $ v$ in the case of dust and $\Lambda=0$. Initial values: $\gamma_{131}=1$, $ v=0.03$, $B=0.03$, $\sigma_{12}=0.03$, $\sigma_{22}=0.03$.}
\label{fig:1} 
\end{figure}

\begin{figure}[h]
\epsfxsize=3in
\epsffile{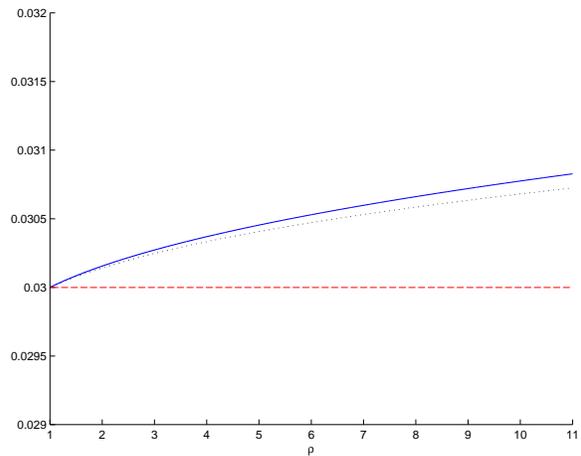}
\vskip4mm
\caption{Comparison between numerical calculation (solid) and perturbations to first (dashed) and second order 
(dotted) for $ v$ in the case of radiation and $\Lambda=0$. Initial values: $\gamma_{131}=1$, $ v=0.03$, $B=0.03$, $\sigma_{12}=0.03$, $\sigma_{22}=0.03$.} 
\label{fig:2}
\end{figure}

To check the perturbative calculation in section \ref{sec:pert} (or conversely the numerical method)
we have solved the system numerically for small deviations (at start) from the Friedmann models
and compared with the perturbative solutions. The constants of integration are chosen so that
the numerical and perturbative solutions agree for the starting value of the independent variable
$\rho$. 
In the two following figures first and second order perturbations together with the numerical solutions for
$ v$ with $\gamma=1$ (fig. \ref{fig:1}) and $\gamma=4/3$ (fig. \ref{fig:2}) are depicted. 
The agreement is similar for the other quantities in $S$ and other values of $\gamma$.

\subsection{Asymptotic behaviour}

\begin{figure} [h]
\hskip1.5cm
\epsfxsize=3in
\epsffile{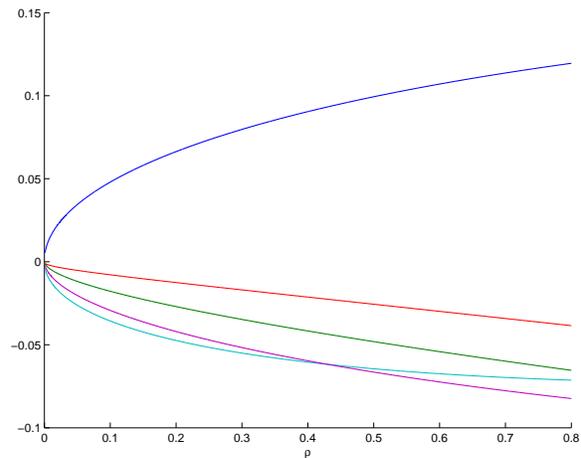} 
\caption{$\sigma_{ij}/\theta\rightarrow0$ as $\rho\rightarrow0$ for dust and $\Lambda=0$. Initial values: $\rho=0.8$, $ v=-0.09$,
$\gamma_{131}=0.08$, $B=0.1$, $\sigma_{12}=0.11$, $\sigma_{22}=0.12$. From top to bottom (at $\rho=0.8$)
$\sigma_{11}/\theta$, $\sigma_{13}/\theta$, $\sigma_{12}/\theta$, $\sigma_{22}/\theta$ and $\sigma_{23}/\theta$} 
\label{fig:3}
\hskip1.5cm
\end{figure}

\begin{figure} [h]
\hskip1.5cm
\epsfxsize=3in
\epsffile{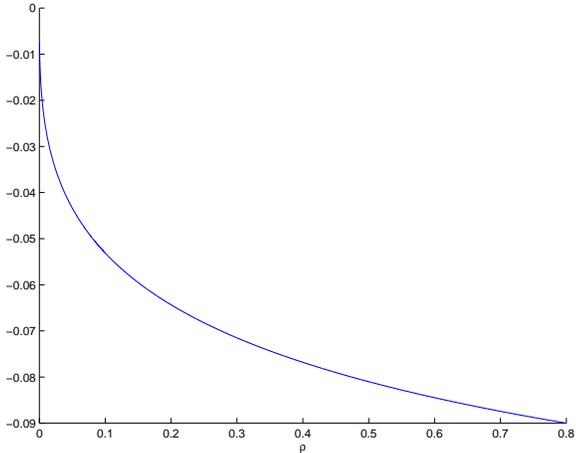} 
\caption{$ v\rightarrow0$ as $\rho\rightarrow0$ for dust and $\Lambda=0$. Initial values as in figure \ref{fig:3}}
\label{fig:4}
\hskip1.5cm
\end{figure}

In \cite{ColeyHervik} a dynamical system analysis was used to find that for $2/3<\gamma<2$ Bianchi type V models approach the Milne universe in the asymptotic future (for low densities).
Figure \ref{fig:3} shows the components of the shear 
normalized with expansion ($\sigma_{ij}/\theta$) for $\gamma=1$, and they 
all vanish in the limit $\rho\rightarrow0$. The tilt $ v$ also approaches zero as shown in figure \ref{fig:4}.

\begin{figure} [h]
\hskip1.5cm
\epsfxsize=3in
\epsffile{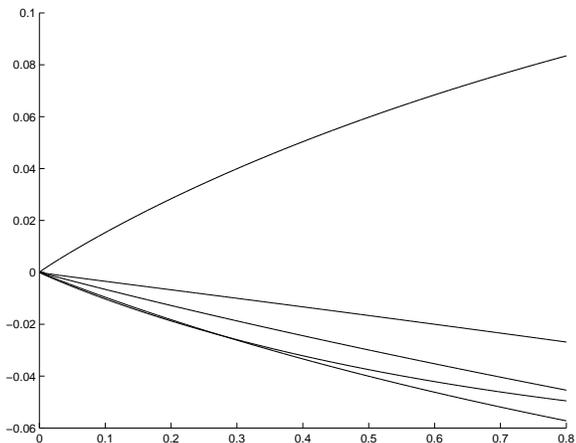} 
\vskip3mm
\caption{$\sigma_{ij}/\theta\rightarrow0$ as $\rho\rightarrow0$ for dust and $\Lambda=1$.
Initial values and ordering same as in figure \ref{fig:3}} 
\label{fig:5}
\hskip1.5cm
\end{figure}

Figure \ref{fig:5} show the corresponding behaviour with a non-zero cosmological constant. 

\begin{figure} [h]
\hskip1.5cm
\epsfxsize=3in
\epsffile{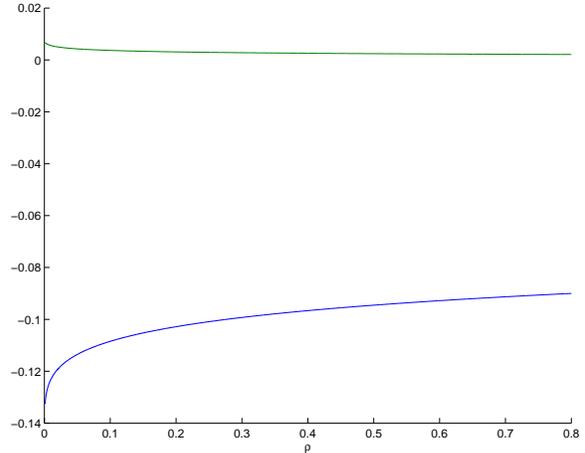} 
\caption{$\omega_{13}/\theta$ (upper curve) and $v$ as $\rho\rightarrow 0$
for radiation and $\Lambda=0$. Initial values: $\rho=0.8$, $ v=-0.09$,
$\gamma_{131}=0.08$, $B=0.1$, $\sigma_{12}=0.11$, $\sigma_{22}=0.12$. }
\label{fig:6}
\hskip1.5cm
\end{figure}

As found for the LRS case in \cite{CollinsEllis} anisotropies may remain for late times with $\gamma \neq 1$.
This is not in conflict with the
result of \cite{ColeyHervik}. An anisotropy remains in the matter fields, but
since matter is getting infinitely diluted spacetime still approaches isotropy.
In figure \ref{fig:6} $\omega_{13}/\theta$ and $ v$ are plotted for $\gamma=4/3$. As seen 
they do not approach zero for small $\rho$. 

\begin{figure} [h]
\hskip1.5cm
\epsfxsize=3in
\epsffile{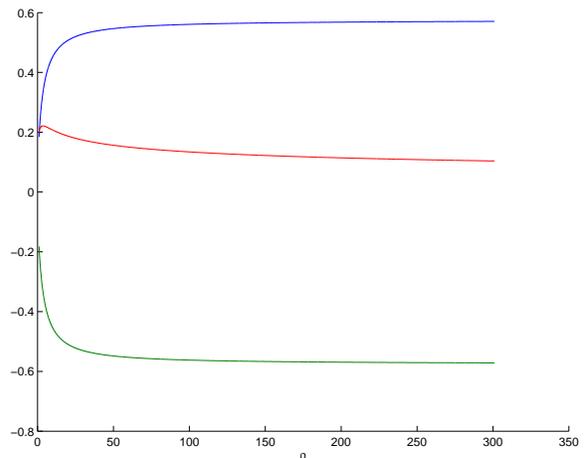} 
\caption{$ v$, $\sigma_{11}/\theta$ and $\sigma_{22}/\theta$ for dust and $\Lambda=0$ when $\rho \rightarrow \infty$.
Initial values: $\rho=1$, $ v=0.2$,
$\gamma_{131}=1$, $B=-0.2$, $\sigma_{12}=0$, $\sigma_{22}=0.7$. 
From top to bottom
$\sigma_{11}/\theta$, $v$ and $\sigma_{22}/\theta$. }
\label{fig:7}
\hskip1.5cm
\end{figure}

\begin{figure} [h]
\hskip1.5cm
\epsfxsize=3in
\epsffile{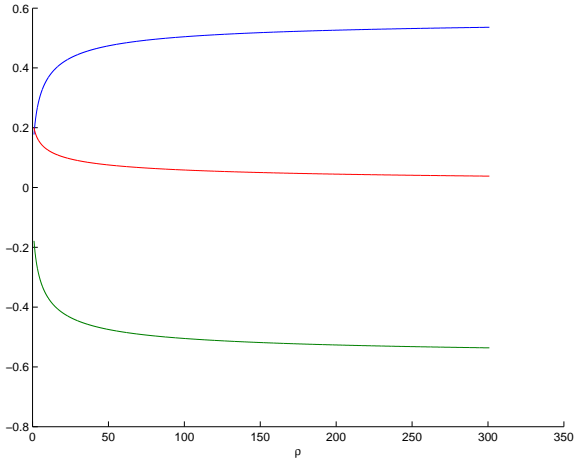} 
\caption{$ v$, $\sigma_{11}/\theta$ and $\sigma_{22}/\theta$ for radiation and $\Lambda=0$ when $\rho \rightarrow \infty$.
Initial values: $\rho=1$, $ v=0.2$,
$\gamma_{131}=1$, $B=-0.2$, $\sigma_{12}=0$, $\sigma_{22}=0.7$. 
From top to bottom
$\sigma_{11}/\theta$, $v$ and $\sigma_{22}/\theta$. }
\label{fig:8}
\hskip1.5cm
\end{figure}

For the LRS case two different behaviours close to the initial singularity was found in \cite{CollinsEllis}. One where
the tilt $ v$ goes to zero and one where it passes the extreme value of one and grows towards infinity. In the second case,
however, often singularities in some of the other quantities occur while $ v$ and the density still are finite. 
The numerical runs seem to show that generically the tilt grows towards one, even if it initially may be decreasing for
a long density interval. In figures \ref{fig:7} and \ref{fig:8} $v$, $\sigma_{11}$ and
$\sigma_{22}$ for $\gamma=1$ and $\gamma=4/3$ are
shown respectively. For these particular cases $v$ was decreasing for as far the numerical calculations could be carried out
(about 20 times as long as shown in the picture). Note that asymptotically $\sigma_{22}=-\sigma_{11}$ and hence these spacetimes do not approach LRS. 

\begin{figure} [h]
\hskip1.5cm
\epsfxsize=3in
\epsffile{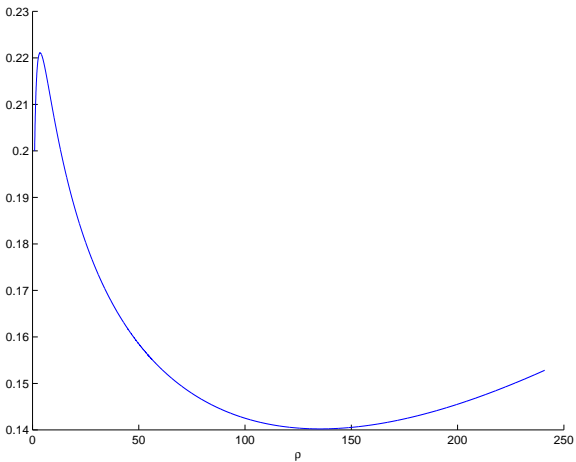} 
\vskip6mm
\caption{$ v$ for dust and $\Lambda=0$ when $\rho \rightarrow \infty$.
Initial values: $\rho=1$, $ v=0.2$,
$\gamma_{131}=1$, $B=-0.2$, $\sigma_{12}=0.01$, $\sigma_{22}=0.7$. }
\label{fig:9}
\hskip1.5cm
\end{figure}

Small changes of the initial values are sufficient to make $v$ eventually turn and start growing as shown in 
 figure \ref{fig:9}. 

Due to an apparent singularity in the equations for $ v=1$ we have not been able to follow the evolution beyond this
value. For the LRS case it is possible to rewrite the only remaining differential equation to avoid this problem, but in the
general case the expressions become quite complex. The equation for $\gamma=4/3$ in the LRS case is given by

\begin{eqnarray}\nonumber
\frac{\dot  v}{ v ^2}=\frac{\left(3 B^2 + 2\rho\right)\left[2 v\left(9B^2+2 v^2\rho\right)\pm\right.}
{2\rho\left(3B^2(9- v^2)+4 v^2\rho\right)
\left(9B^2+( v^2+3)\rho\right)}\\\label{betarad}
\left. \pm \left( v^2-3\right)\sqrt{81B^4+9B^2 v^2\rho+27B^2\rho+4 v^2\rho^2}\right]\, ,
\end{eqnarray}
where $B$ is given by (\ref{Bbeta}). For the negative root the tilt may grow larger than one as seen in figure \ref{fig:10}:

\begin{figure} [h]
\hskip1.5cm
\epsfxsize=3in
\epsffile{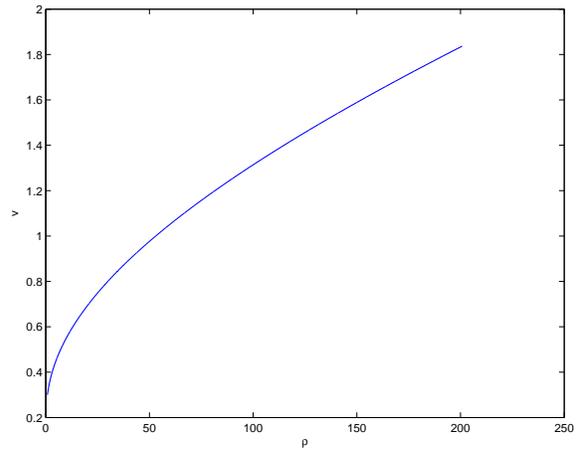} 
\vskip5mm
\caption{$ v$ as a function of $\rho$ for radiation and $\Lambda=0$ with the negative root in (\ref{betarad}). Initial values: $\rho=0.8$, $ v=0.3$. $C_1=1$.}
\label{fig:10}
\hskip1.5cm
\end{figure}

For this case, however, the kinematic quantities as well as the Weyl tensor diverge for $ v=\sqrt{3}$, as also found in \cite{CollinsEllis}.

\section{Conclusions}

In this paper it was shown that the general Bianchi V cosmology with linear equation of state and
cosmological constant can be reduced to
an integrable system of five ordinary first order differential equations for quantities that give a complete
local description of the geometry. The full line-element was found in terms of quadratures of these quantities.
In general the solutions have expansion, shear and vorticity. 
The system was cast in a form suitable for perturbative calculations
and the first order perturbations around the open Friedmann model with vorticity, being approximations to exact
solutions, were constructed. Perturbative calculations to higher orders would be straightforward up to quadratures.

A numerical study was done and the results agree well with the perturbative ones in the appropriate domains.
For large times (small densities) the results agree well with previous works \cite{CollinsEllis,HewittWainwright,ColeyHervik}.
Numerically we found that the tilt probably falls off towards zero when the density grows for special initial values, but generically
it seems as if the tilt eventually always grows unlimited for large densities. In \cite{HewittWainwright,ColeyHervik} the late time behaviour
of (among others) Bianchi V solutions were studied using dynamical systems analysis. It would be of great interest to extend this
analysis to early times.

\acknowledgments

We are grateful to Istv\'an R\'acz and Krister Wiklund for helpful discussions.

\end{document}